\documentclass{article}
\usepackage{graphicx}
\usepackage{epsfig}
\usepackage{url}
\usepackage{psfrag}
\usepackage{times}
\usepackage{amsfonts,amssymb,amsbsy}
\usepackage{amsmath}
\usepackage{latexsym}
\usepackage{amsfonts,amsbsy}

\title{Rumsey's Reaction Concept Generalized}
\author{I.V. Lindell and A. Sihvola} 
\date{School of Electrical Engineering,\\ Aalto University, Espoo, Finland\\ 
{\tt ismo.lindell@aalto.fi}\\\vspace{-1pt}{\tt ari.sihvola@aalto.fi}}
\parskip=\medskipamount

\begin{document}


\def\e{\begin{equation}} 
\def\f{\end{equation}} 
\def\ea{\begin{eqnarray}} 
\def\fa{\end{eqnarray}} 

\def\##1{{\mbox{\textbf{#1}}}}
\def\%#1{{\mbox{\boldmath $#1$}}}
\def\=#1{{\overline{\overline{\mathsf #1}}}}
\def\RR{\mbox{\boldmath $\R$}}
\def\nn#1{{\sf #1}}
\def\SE{{\mathbb E}}
\def\SF{{\mathbb F}}

\def\*{^{\displaystyle*}}
\def\xx{\displaystyle{{}^\times}\llap{${}_\times$}}
\def\.{\cdot}
\def\x{\times}
\def\oo{\infty}

\def\D{\nabla}
\def\d{\partial}

\def\ra{\rightarrow}
\def\lra{\leftrightarrow}
\def\Ra{\Rightarrow}
\def\le{\left(}
\def\ri{\right)}
\def\l#1{\label{eq:#1}}
\def\r#1{(\ref{eq:#1})}
\def\am{\left(\begin{array}{c}}
\def\amm{\left(\begin{array}{cc}}
\def\ammm{\left(\begin{array}{ccc}}
\def\ammmm{\left(\begin{array}{cccc}}
\def\a{\end{array}\right)}

\def\I{\int\limits}
\def\OI{\oint\limits}

\def\A{\alpha}
\def\B{\beta}
\def\de{\delta}
\def\De{\Delta}
\def\E{\epsilon}
\def\g{\gamma}
\def\G{\Gamma}
\def\h{\eta}
\def\K{\kappa}
\def\la{\lambda}
\def\La{\Lambda}
\def\M{\mu}
\def\o{\omega}
\def\Om{\Omega}
\def\R{\rho}
\def\s{\sigma}
\def\t{\tau}
\def\z{\zeta}
\def\X{\chi}
\def\TH{\theta}
\def\Th{\Theta}
\def\VF{\varphi}
\def\VR{\varrho}
\def\VT{\vartheta}
\def\ve{\%\varepsilon}

\def\tr{{\rm tr }}
\def\spm{{\rm spm}}
\def\det{{\rm det}}
\def\Det{{\rm Det}}
\def\sgn{{\rm sgn}}
\def\bi{\bibitem}

\def\W{\wedge}
\def\WW{\displaystyle{{}^\wedge}\llap{${}_\wedge$}}
\def\Adj{{\rm Adj\mit}}
\def\ua{\uparrow}
\def\da{\downarrow}
\def\uda{\updownarrow}

\def\J{\rfloor}
\def\L{\lfloor}
\def\JJ{\rfloor\rfloor}
\def\LL{\lfloor\lfloor}

\maketitle

\begin{abstract}
The reaction concept, introduced by Rumsey in 1954, describes interaction between time-harmonic electromagnetic sources through the fields radiated by the sources. In the original form the concept was a scalar quantity defined by three-dimensional field and source vectors. In the present paper, the representation is extended to four dimensions applying differential-form formalism. It turns out that, in a coordinate-free form, the reaction concept must actually be a one-form, whose temporal component yields Rumsey's scalar reaction. The spatial one-form component corresponds to a three-dimensional Gibbsian-vector reaction which consists of electromagnetic force terms.  
\end{abstract}



\section{Introduction}

The concept of reaction between two electromagnetic sources was introduced by V.H. Rumsey in 1954 as "a physical observable like mass, length, charge, etc." \cite{Rumsey54,Rumsey59}. Assuming two sets of monochromatic time-harmonic electric and magnetic current sources $\#J_{eg}^a,\#J_{mg}^a$ and $\#J_{eg}^b,\#J_{mg}^b$, the reaction of sources b on the sources a through the fields $\#E_g^b,\#H_g^b$ created by the sources b is defined by 
\e <ab> = \I_{V_a} R^{ab}dV, \l{ab}\f
where $R^{ab}$ is the reaction density,
\e R^{ab}= \#J_{eg}^a\.\#E_g^b - \#J_{mg}^a\.\#H_g^b. \l{Rab} \f
The subscript $()_g$ is added to emphasize the 3D Gibbsian vector character of the quantities \cite{Gibbs}, to distinguish them from the 4D differential-form representations of the same physical quantities discussed in the subsequent Section. 

The integration in \r{ab} is over a finite region $V_a$ which contains the sources a and excludes the sources b. The minus sign between the terms in \r{ab} can be justified by the minus sign in the Maxwell equations, see Appendix for a clarification. 

The system is reciprocal when the condition
\e <ab> = <ba> \l{abba}\f
is valid \cite{Rumsey54}. When the sources a and b are in different media, the reciprocity principle must be taken in mdified form \cite{Kong72,Altman,Kong}.

Obviously, the reaction \r{ab} is a scalar quantity. In \cite{Rumsey54}, the reaction involving electric charges $\VR_e$ as the sources was defined by
\e (ab) = \I_{V_a} \#E_g^b \VR_e^a dV.  \l{(ab)} \f
In this case, the reaction is a vector quantity: the force exerted by the electric field b on the electric charge a. 

Over the years following its introduction, the reaction concept \r{ab} has found application in solving electromagnetic problems. For example, impedance parameters of multiport networks, resonant frequencies of cavities, cut-off frequencies of waveguides, input impedances of antennas and scattering cross sections of obstacles could be shown to be proportional to reaction quantities, which helped finding simple numerical solutions to practical problems \cite{Harrington,Moment,Cohen55,Richmond61,Balanis,Wang75}. 

It would be interesting to generalize the reaction concept so that both \r{ab} and \r{(ab)} would fall under the same definition. For this we need a tour through the 4D formalism. Previously, the reaction concept \r{ab} has been generalized to sources of more general time-dependence \cite{Welch60,Bojarsky83}. However, here we assume time-harmonic sources and fields. Also, the medium is assumed isotropic with parameters $\E_o,\M_o$, for simplicity.

\section{4D Representation of Quantities}

Assuming a 3D vector basis $\#e_1,\#e_2,\#e_3$, let us expand the Gibbsian field and source vectors as
\ea \#E_g &=& \#e_1 E_1+ \#e_2 E_2+ \#e_3 E_3, \\
\#H_g &=& \#e_1 H_1+ \#e_2 H_2 + \#e_3 H_3, \\
\#J_{eg} &=& \#e_1 J_{e23}+ \#e_2 J_{e31}+ \#e_3 J_{e12}, \\
 \#J_{mg} &=& \#e_1 J_{m23}+ \#e_2 J_{m31}+ \#e_3 J_{m12}. \fa
The two scalar quantities appearing in the definition of the reaction density \r{Rab} have the expansions
\ea \#J_{eg}\.\#E_g &=& E_1 J_{e23}+ E_2 J_{e31}+ E_3 J_{e12}, \l{fe}\\
    \#J_{mg}\.\#H_g &=& H_1 J_{m23}+ H_2 J_{m31} + H_3 J_{m12}. \l{fm}\fa

Applying the 4D formalism, the spatial 3D vector basis will be extended by a temporal vector $\#e_4$. The electric and magnetic Gibbsian field vectors $\#E_g,\#H_g$ are represented by field one-forms $\#E,\#H$. Details of the formalism applied here can be found in \cite{Difform,MDEM}. A spatial basis of one-forms $\ve_1, \ve_2, \ve_3$ with the temporal one-form $\ve_4$ is chosen dual to the basis of vectors $\#e_i$ as to satisfy $\#e_i|\ve_j=\de_{ij}$. The field one-forms can be expanded as
\ea \#E &=& \ve_1E_1+ \ve_2E_2+ \ve_3E_3 \\
&=& \=\G_s|(\#e_1E_1+\#e_2E_2+ \#e_3E_3) \\
&=& \=\G_s|\#E_g, \\
\#H &=& \ve_1H_1+ \ve_2H_2+ \ve_3H_3 \\
&=& \=\G_s|(\#e_1H_1+\#e_2H_2+ \#e_3H_3) \\
&=& \=\G_s|\#H_g, \fa
where $\=\G_s$ is the spatial metric dyadic \cite{Difform},
\e \=\G_s = \ve_1\ve_1+ \ve_2\ve_2 + \ve_3\ve_3, \f
mapping spatial vectors to spatial one-forms. Its spatial inverse 
\e \=G_s = \#e_1\#e_1+ \#e_2\#e_2+ \#e_3\#e_3, \f
maps one-forms to vectors.

The Gibbsian source vectors $\#J_{eg},\#J_{mg}$ are represented by source two-forms $\#J_e,\#J_m$ defined by
\ea \#J_e &=& \ve_{12} J_{e12} + \ve_{23} J_{e23} + \ve_{31} J_{e31} \\
&=&  \ve_{123}\L(\#e_1 J_{e23}+ \#e_2 J_{e31}+ \#e_3 J_{e12}) \\
&=& \ve_{123}\L\#J_{eg},\\
\#J_m &=& \ve_{12} J_{m12} + \ve_{23} J_{m23} + \ve_{31} J_{m31} \\
&=&  \ve_{123}\L(\#e_1 J_{m23}+ \#e_2 J_{m31}+ \#e_3 J_{m12}) \\
&=& \ve_{123}\L\#J_{mg}.\fa
The wedge products of basis one-forms $\ve_{ij}=\ve_i\W\ve_j$ make a basis of two-forms. $\ve_{123}=\ve_1\W\ve_2\W\ve_3$ is the spatial three-form and $\ve_N=\ve_{1234}$ is the basis four-form. Basis bivectors are defined by $\#e_{ij}=\#e_i\W\#e_j$, trivectors by $\#e_{ijk}$ and $\#e_N=\#e_{1234}$ is the quadrivector formed by the basis vectors. $\L$ is the contraction operation satisfying
\e \ve_{123}\L\#e_1=\ve_{23},\ \ \ve_{123}\L\#e_2=\ve_{31},\ \ \ \ve_{123}\L\#e_3=\ve_{12},\f
\e \ve_N\L\#e_1 = \ve_{234},\ \ \ve_N\L\#e_2= \ve_{314},\f
\e \ve_N\L\#e_3= \ve_{124},\ \ \ve_N\L\#e_4=-\ve_{123}. \f

Applying 
\ea \#J_{eg}\.\#E_g &=&  \#J_{eg}|\#E = (\#e_{123}\L\#J_e)|\#E \\
&=& \#e_{123}|(\#J_e\W\#E)=  \#e_N|(\#J_e\W\#E\W\ve_4), \\
 \#J_{mg}\.\#H_g &=& \#J_{mg}|\#H = (\#e_{123}\L\#J_m)|\#H \\
 &=& \#e_{123}|(\#J_m\W\#H) = \#e_N|(\#J_m\W\#H\W\ve_4), \fa
the reaction density \r{Rab} can be expressed in terms of 4D quantities as
\e R^{ab} = \#e_N|(\#J_e^a\W\#E^b\W\ve_4  - \#J_m^a\W\#H^b\W\ve_4). \l{Rab1}\f

\section{Extending the Reaction Concept}

Let us further express the reaction density \r{Rab1} in terms of more general 4D field and source quantities. The basic electromagnetic two-forms are defined by \cite{Difform}
\ea \%\Phi &=& \#B +\#E\W\ve_4,\l{Phi}\\
    \%\Psi &=& \#D -\#H\W\ve_4, \l{Psi}\fa
where $\#B$ and $\#D$ are spatial field two-forms. Since $\#J_e\W\#B$ and $\#J_m\W\#D$ are spatial four-forms, they actually vanish, whence \r{Rab1} takes the form
\ea R^{ab} &=& \#e_N|(\#J_e^a\W\%\Phi^b+\#J_m^a\W\%\Psi^b) \\
&=& \#e_N|(\%\Phi^b\W\#J_e^a+\%\Psi^b\W\#J_m^a). \l{fab1}\fa
The electric and magnetic source three-forms are defined by \cite{Difform}
\ea \%\g_e &=& \%\VR_e - \#J_e\W\ve_4,\l{ge}\\
    \%\g_m &=& \%\VR_m - \#J_m\W\ve_4, \l{gm}\fa
where $\%\VR_e=\VR_e\ve_{123}$ and $\%\VR_m=\VR_m\ve_{123}$ denote electric and magnetic charge three-forms. The source two-forms $\#J_e$ and $\#J_m$ can be obtained from the corresponding three-forms through contraction as
 \ea \#J_e &=& -\%\g_e\L\#e_4, \\
     \#J_m &=& -\%\g_m\L\#e_4. \fa
Substituting these in \r{Rab1} yields the expression
\e R^{ab} = -\#e_N|(\%\Phi^b\W(\%\g_e^a\L\#e_4)+\%\Psi^b\W(\%\g_m^a\L\#e_4)). \l{Rab2}\f
It is desirable to find a representation which is independent of the chosen basis. Obviously, \r{Rab2} depends on $\#e_4$, chosen to represent the temporal basis vector. Multiplying the expression by the corresponding temporal one-form $\ve_4$ as
\e R^{ab}\ve_4 = -\#e_N|(\%\Phi^b\W(\%\g_e^a\L\#e_4\ve_4)+\%\Psi^b\W(\%\g_m^a\L\#e_4\ve_4)), \l{Rab3}\f 
and replacing the dyadic product $\#e_4\ve_4$ by the unit dyadic $\=I=\sum \#e_i\ve_i$, the scalar quantity $R_{ab}$ gives rise to the coordinate-independent one-form 
\e \#R^{ab} = -\#e_N|(\%\Phi^b\W(\%\g_e^a\L\=I)+\%\Psi^b\W(\%\g_m^a\L\=I)), \l{Rab4}\f 
which is equivalent with
\e \#R^{ab} = -(\#e_N\L\%\Phi^b)\L\%\g_e^a - (\#e_N\L\%\Psi^b)\L\%\g_m^a. \f
Because the scalar \r{Rab} can be obtained as the temporal component of the one-form \r{Rab4}, 
\e R^{ab} = \#R^{ab}|\#e_4, \f
$\#R^{ab}$ can be conceived as a generalization of the scalar reaction density $R^{ab}$. Because all quadrivetors are multiples of one another, the basis quadrivector $\#e_N$ could be replaced by any other quadrivector, whence the reaction density quantity is actually nonunique. However, the scalar factor cancels out in the reciprocity rule \r{abba}.

\section{Spatial Component of Reaction Density One-Form}

Since the temporal component of the extended reaction density \r{Rab4} yields the classical reaction density, it is interesting to study more closely its spatial component. Denoting the spatial unit dyadic by
\e \=I_s = \#e_1\ve_1+ \#e_2\ve_2+ \#e_3\ve_3, \f
the spatial component of the generalized reaction density \r{Rab4} is defined by
\e \#R_s^{ab} = \#R^{ab}|\=I_s = -\#e_N|(\%\Phi^b\W(\%\g_e^a\L\=I{}_s)+\%\Psi^b\W(\%\g_m^a\L\=I_s)). \l{Rabs}\f

Applying
\ea \%\g_e\L\=I_s &=& (\%\VR_e - \#J_e\W\ve_4)\L\=I_s \\
&=& \VR_e\ve_{123}\L\=I_s - \ve_4\W(\#J_e\L\=I_s), \fa
we can expand the first term of \r{Rabs} as 
\ea  -\#e_N|(\%\Phi\W(\%\g_e\L\=I_s)) &=& -\#e_N|(\%\Phi\W(\VR_e\ve_{123}\L\=I_s- \ve_4\W(\#J_e\L\=I_s)))\\ 
&=& -\#e_N|(\VR_e(\#E\W\ve_4)\W(\ve_{123}\L\=I_s) - \#B\W\ve_4\W(\#J_e\L\=I_s))\\ 
&=& -\VR_e\#e_{123}|(\#E\W(\ve_{123}\L\=I_s)) - \#e_{123}|(\#B\W(\#J_e\L\=I_s)) \\
&=& -\VR_e\#E - (\#e_{123}\L\#B)|\W(\#J_e\L\=I_s), \\ 
&=& -\VR_e\#E  -(\#B_g\x\#J_{eg})|\=\G_s. \l{fabs}\fa 
The last term of \r{fabs} can be verified by expanding the two-forms $\#B$ and $\#J_e$ in their components. The second term of \r{Rabs} can be expanded similarly as 
\e -\#e_N|(\%\Psi\W(\%\g_m\L\=I_s)) =\VR_m\#H -(\#D_g\x\#J_{mg})|\=\G_s. \l{Psigm}\f
Actually, this result can be directly written from \r{fabs} by changing the symbols as $\%\Phi\ra\%\Psi$, $\%\g_e\ra\%\g_m$, which, from \r{Phi} and \r{Psi}, together with \r{ge} and \r{gm}, implies $\#B\ra\#D$, $\#E\ra-\#H$, $\s_e\ra\s_m$ and $\#J_e\ra\#J_m$, and, similarly, for the corresponding Gibbsian vector symbols \cite{Liu}.

Combining the expressions, the spatial part of the extended one-form reaction density one-form \r{Rab4} corresponds to the Gibbsian vector quantity
\e \#R_{sg}^{ab} = \#R_s^{ab}|\=G_s = -\VR_e^a\#E_g^b + \VR_m^a\#H_g^b - \#B_g^b\x\#J_{eg}^a - \#D_g^b\x\#J_{mg}^a, \l{Rsgab}\f
in terms of which the total reaction density vector can be expressed as
\e\#R_g^{ab} =  (\#J_{eg}^a\.\#E_g^b - \#J_{mg}^a\.\#H_g^b)\#e_4 -\VR_e^a\#E_g^b + \VR_m^a\#H_g^b - \#B_g^b\x\#J_{eg}^a - \#D_g^b\x\#J_{mg}^a, \l{Rgab}\f
which is a generalization of the scalar quantity \r{Rab}. The spatial vector consists of terms corresponding to forces on electric and magnetic charges and Lorentz forces on electric and magnetic currents.

\section{Conclusion}

The classical reaction concept, introduced by V.H. Rumsey in 1954, has been generalized in four-dimensional formalism from a scalar quantity to a one-form quantity. This corresponds in Gibbsian three-dimensional formalism to a combination of scalar and vector components. The reaction one-form is independent of the choice of temporal basis  one-form. The novel Gibbsian vector component consists of force terms on electric and magnetic charge and current sources. In this analysis, the medium is assumed isotropic, and the sources and fields are assumed to have monochromatic time-harmonic time dependence.

\section*{Appendix: Justification of Rumsey's Expression}

To justify the minus sign in Rumsey's expression for the reaction \r{ab}, let us assume that the reaction of an Gibbsian vector electric source $\#J_{eg}^a$ is of the form
\e <ab> = \I_{V_a} \#E_g^b\.\#J_{eg}^a dV. \f
Let us add another electric source, which is actually a magnetic source $\#J_m^a$, which in an isotropic medium can be represented by the equivalent electric source \cite{Mayes,Methods}
\e \#J_{eqg}^a = \frac{1}{j\o\M}\D\x\#J_{mg}^a. \f 
In this case, the reaction  can be expressed as
\ea <ab> &=& \I_{V_a} \#E_g^b\.(\#J_{eg}^a + \#J_{eqg}^a) dV  \\
&=& \I_{V_a} \#E_g^b\.(\#J_{eg}^a + \frac{1}{j\o\M_o}\D\x\#J_{mg}^a) dV \\
&=& \I_{V_a} (\#E_g^b\.\#J_{eg}^a - \frac{1}{j\o\M_o}(\D\.(\#E_g^b\x\#J_{mg}^a)+ j\o\M_o\#H_g^b\.\#J_{mg}^a) dV. \fa
The divergence term vanishes when the volume of integration has a boundary outside the sources a. In this case the expression reduces to that of \r{ab}, thus justifying the minus sign between the two terms.

\section*{Acknowledgment} 
The authors thank Drs Guoqiang Liu, Jing Liu and Yuanyuan Li, from Institute of Electrical Engineering, Chinese Academy of Sciences, School of Electronic, Electrical and Communications Engineering, University of Chinese Academy of Sciences, China, for their help in finding the correct form of Equations \r{Psigm} -- \r{Rgab}, \cite{Liu}.

\end{document}